\documentclass[preprint,12pt]{elsarticle}
\usepackage[psamsfonts]{amssymb}
\usepackage{amsmath}

\newcommand{\be}{\begin{equation}}
\newcommand{\ee}{\end{equation}}
\newcommand{\bea}{\begin{eqnarray}}
\newcommand{\eea}{\end{eqnarray}}

\newcommand{\n}{\nonumber}

\begin{document}

\begin{frontmatter}
\title{Nonlinear Coherent States of the Fokas-Lagerstrom Potential}
\author{M. Ashrafi $^1$}

\author{A. Mahdifar $^{2,3}$}

\ead{a.mahdifar@sci.ui.ac.ir}

\author{E. Amooghorban $^{1,4}$}

\address{ $^1$ Department of Physics, Faculty of Basic Sciences, Shahrekord University, P.O. Box 115, Shahrekord 88186-34141, Iran.}
\address{$^2$ Department of Physics, Faculty of Science, University of Isfahan, Hezar Jerib, Isfahan, 81746-73441,
         Iran }

\address{$^3$ Quantum Optics Group, Department of Physics, Faculty of Science, University of Isfahan, Hezar Jerib Str., Isfahan, 81746-73441, Iran}

\address{$^4$ Nanotechnology Research Center, Shahrekord University, Shahrekord, 88186-34141, Iran.}

\begin{abstract}
In this paper, we introduce an algebraic approach to construct Fokas-Lagerstrom coherent states.
To do so, we define deformed creation and annihilation
operators associated to this system and investigate their algebra.
We show that these operators satisfy the $f$-deformed Weyl-Heisenberg algebra.
Then, we propose a theoretical scheme to generate the aforementioned coherent states.
The present contribution shows that the Fokas-Lagerstrom nonlinear coherent states possess some
non-classical features.
\end{abstract}


\end{frontmatter}

\section{Introduction}

%Deformed oscillator algebra
It is well known that an $N$-dimensional classical or quantum system is called completely integrable if
there are $N$ functionally independent well defined constants of motion (including the Hamiltonian) in involution \cite{Arnold, Winternitz}.
Furthermore, if it is possible to obtain $2N-1$ constants
of motion, the system is called maximally superintegrable, provided that the Poisson brackets (or the commutators) of
these constants of motion with the Hamiltonian vanish (in general, only $N$ of $2N-1$ constants are in involution) \cite{Winternitz,5,2,3,4,6,7}.
Famous examples of the superintegrable systems are the isotropic harmonic oscillator and the Kepler-Coulomb potential.
It is worthwhile to note that their higher order symmetries, play the fundamental role in solvability and other
interesting properties of these physical systems.

In the last decades, coherent states of the harmonic oscillator \cite{Glauber} and
generalized coherent states associated with various algebras \cite{Perelomov, Klauder, Twareqe} have
been playing an important role in various branches of physics.
The coherent states, which defined as the right eigenstates of the annihilation
operator $\hat{a}$, are the quantum states with classical-like properties. In other words,
they are the closest analogue to classical states.
On the other hand, the generalized coherent states exhibit
some nonclassical properties and, therefore, have received an ever-increasing interest during the last decades.
Among the generalized coherent states, the nonlinear coherent states or f-deformed coherent states \cite{Solomon, Rokni, Rokni2} have
attracted many interests in recent years due to their applications in quantum optics
and quantum technology.
It is shown that the statistical characteristics of these states exhibit some
nonclassical features, such as photon antibunching \cite{Kimble},
sub-Poissonian photon statistics \cite{Teich} and squeezing \cite{Hong}. These states, which are
associated with nonlinear algebras \cite{Solomon, Manko}, could be generated in the
center-of-mass motion of an appropriately laser-driven trapped ion \cite{Vogel, Mahdifar2008} and
in a micromaser under intensity-dependent atom-field interaction \cite{Naderi}.

Recently, one of us defined the nonlinear coherent states for some of the two-dimensional
superintegrable systems, including the isotropic harmonic oscillator on a sphere \cite{Mahdifar2006} and the
Kepler-Coulomb problem on a sphere \cite{Hoseinzadeh}.
In the present contribution, we study the Fokas-Lagerstrom potential as the another example of the two-dimensional
superintegrable systems \cite{Fokas}.
Our approach which is based on the f-deformed harmonic oscillator algebra and the nonlinear coherent states,
can increase the insight about the Fokas-Lagerstrom system. Of course, the Fokas-Lagerstrom potential was considered previously \cite{Fokas, Bonatsos}.
The distinction between the present paper and the mentioned references is that our approach is based on the algebraic
methods of the nonlinear coherent states.

The paper is organized as follows.
In Sec. \ref{Suint}, we briefly review the superintegrable systems.
By using the nonlinear oscillator approach in Sec. \ref{NLApp}, we investigate the Fokas-Lagerstrom system, as an
example of two-dimensional superintegrable systems and show that the algebra of the this system can be considered as
an $f$-deformed Weyl-Heisenberg algebra.
We define nonlinear coherent states for the Fokas-Lagerstrom potential and examine their resolution of identity in Sec.
\ref{FLCS}.
In Sec.~\ref{Gen}, we propose a scheme to generate the  aforementioned coherent states.
Sec.~\ref{QStatistical} is devoted to the study of the quantum statistical properties of the
constructed nonlinear coherent states, including mean number of photons, Mandel parameter and
quadrature squeezing. Finally, the summary
and concluding remarks are given in Sec.~\ref{Summary}.

\section{Superintegrable systems}\label{Suint}

In classical mechanics, an $N$-dimensional system is called superintegrable, if it has more than $N$ independent
constants of motion. Furthermore, if the system has exactly $(2N-1)$ independent constants of motion (the maximum number)
and also if all of these constants are single valued and globally defined, then the system is called maximally
superintegrable system \cite{Arnold, Winternitz}.

To be specific, for a classical system in two dimensions, described by the following Hamiltonian:
\begin{equation}\label{1}
    H = H\left( {x,y,{p_x},{p_y}} \right),
\end{equation}
if there exist two independent constants of motion $C_{1}$ and $C_{2}$, so that we have,
\begin{eqnarray}\label{2}
    && {\left\{ {H,\left. C_{1} \right\}} \right._{PB}} = {\left\{ {H,\left. C_{2} \right\}} \right._{PB}} = 0,\nonumber\\
    && {\left\{ {C_{1},\left. C_{2} \right\}} \right._{PB}} \ne  0,
\end{eqnarray}
then this system is  called a superintegrable system.
Here, $\{ \;,\;\}_{PB}$ denotes the Poisson bracket~\cite{Winternitz}.

In quantum mechanics, a two-dimensional system described by a Hamiltonian $\hat H$ is called integrable, if is possible to find an operator $\hat C_{1}$ commuting with the $\hat H$ ~\cite{Winternitz}:
     \begin{equation}\label{3}
     [\hat H,\hat C_{1}] = 0.
     \end{equation}
This system is also called superintegrable, if there exists another operator $\hat C_{2}$, linearly independent of $\hat H$
and $\hat C_{1}$, that commute with $\hat H$ but not commute with $\hat C_{1}$, so that,
\begin{equation}\label{4}
     [\hat H,\hat C_{2}] = 0,\\
     \ [\hat C_{1},\hat C_{2}] \ne 0.
\end{equation}

\section{Fokas- Lagerstrom system}\label{NLApp}

We consider the quantum Fokas-Lagerstrom system, described by the Hamiltonian~\cite{Fokas}:
\begin{equation}\label{5}
\hat H = \frac{1}{2}(\hat P_x^2 + \hat P_y^2) + \frac{{{{\hat x}^2}}}{2} + \frac{{{{\hat y}^2}}}{{18}},
\end{equation}
(in this paper we put $\hbar = m = \omega  = 1$).
By introducing the following operators:
\begin{eqnarray}\label{6}
      \hat J &=&\hat P_x^2 +\hat {x^2},\\
      \hat B &=& \frac{1}{2}\{ \hat X{{\hat P}_y} - \hat Y{{\hat P}_x},\hat P_y^2\}  + \frac{{{{\hat Y}^3}{{\hat
      P}_x}}}{{27}} - \frac{{\{       \hat X{{\hat Y}^2},{{\hat P}_y}\} }}{6},
\end{eqnarray}
where $\{ , \}$ is the usual anticommutator, and making use of Eq.(\ref{5}), it can be easily shown that the following commutation relations hold~\cite{Bonatsos},
\begin{eqnarray}\label{7}
        \left[\hat H,\hat B \right] &=&\left[\hat H,\hat J \right]= 0,\nonumber\\
        \left[ {\hat J,\hat R} \right] &=& 4\hat B,\\
        \left[ {\hat J,\hat B} \right] &=& \hat R.\nonumber
\end{eqnarray}
With these in mind, and considering the definition of the superintegrable systems, it is clear that the Fokas-Lagerstrom system is a quantum superintegrable system. Now, by defining the following operators:
\begin{eqnarray}\label{8}
         \hat n &=& \frac{{\hat J}}{2} - u\hat I,\n\\
         {\hat A}^\dag &=& \hat B + \frac{\hat R}{2},\n\\
         \hat A &=& \hat B - \frac{\hat R}{2},
\end{eqnarray}
where $u$ is a constant to be determined, it is seen that the operators $\hat A$, ${\hat A}^\dag $ and $\hat n$ satisfy the following closed algebra~\cite{Bonatsos}:
\begin{eqnarray}\label{9}
         \ [\hat n,{\hat A}^\dag]&=&\hat{A}^{\dag},\n\\
         \ [\hat n,{\hat A}]&=&-\hat{A},\n\\
         \ [\hat A,{\hat A^\dag }] &=& \Phi (\hat H,\hat{n} + 1) - \Phi (\hat H,\hat{n}).
\end{eqnarray}
where the structure function $\phi$ is given by
\begin{eqnarray}\label{10}
         \Phi (E,x) &=& \frac{1}{9}(2x + 2u - 1)(2E + 1 - 2x - 2u)\n\\
          &&\times (6E - 6u - 6x)(6E - 6u + 5 - 6x).
\end{eqnarray}
This structure function, for $n\geq0$ is a real definite positive function and also $\phi \left( {E,0} \right) = 0$.
Now, for each energy eigenvalue it is possible to define the corresponding Fock space as below \cite{Bonatsos, Panahi},
\begin{eqnarray}\label{11}
         \hat H\left| {E,n} \right\rangle &=& E\left| {E,n} \right\rangle ,\nonumber\\
         \hat N\left| {E,n} \right\rangle &=& n\left| {E,n} \right\rangle , \nonumber\\
         \hat A\left| {E,0} \right\rangle &=& 0, \nonumber\\
         \left| {E,n} \right\rangle &=& \frac{\left(
         {{A^\dag }} \right)^n}{{\sqrt {\left[ {\phi (E,n)} \right]} !\,}}  \left| {E,0}               \right\rangle .
\end{eqnarray}
For a discrete energy eigenvalue $E$, there is the $N+1$-dimensional degeneracy. So, we deal with an $N$-dimensional Fock space corresponding to that eigenvalue energy as,
\begin{eqnarray}\label{12}
         \hat H\left| {N,n} \right\rangle  &= &{E_N}\left| {N,n} \right\rangle ,\quad \quad \quad \quad N = 0,1,2,...,\\
         \hat n\left| {N,n} \right\rangle & =& n\left| {N,n} \right\rangle \quad \quad \quad \quad \quad  n=0,1,2,...,N.
\end{eqnarray}
The combination of this restriction along with $ \phi \left( {E,0} \right) = 0$ and  $ \phi \left( {E,N+1} \right) = 0$,
determines $u$ and the possible energy eigenvalues as below \cite{Bonatsos}:
\begin{eqnarray}\label{13}
       u     &=& \dfrac{1}{2},\\
       {E_N} &=& N + 1,\\
       {E_N} &=& N + \frac{2}{3},\\
       {E_N} &=& N + \frac{4}{3}.
\end{eqnarray}
Therefore, the allowable structure functions are respectively given by,
\begin{eqnarray}\label{14}
       \Phi ({E_N},x) = 16x(N + 1 - x)(N + A - x)(N + B - x),
\end{eqnarray}
where, the constants $A$ and $B$ are corresponding to one of the pairs $(2/3,4/3)$, $(2/3,1/3)$ or $(5/3,4/3)$.\\
As is well known, the $f$-deformed annihilation and creation operators associated with an $f$-deformed
harmonic oscillator can be defined as ~\cite{Manko},
\begin{eqnarray}\label{15}
       \hat{A} &=&\hat{a} f (\hat{n}) =f ( \hat{n}+1) \hat{a},\n\\
       \hat{A}^{\dagger} &=& f^{\dagger} (\hat{n}) \hat{a}^{\dagger}= \hat{a}^{\dagger} f^{\dagger} (\hat{n}+1),
\end{eqnarray}
where $ \hat a$, $ \hat{a}^{\dagger}$ and $ \hat n$ are the bosonic annihilation, creation and number operators,
respectively, and $f(\hat{n})$ is a real nonnegative deformation function. These deformed operators satisfy the commutation relation
\begin{eqnarray}\label{16}
       [\hat A,{\hat A^\dag }] = (\hat n + 1)f(\hat n + 1){\hat f^\dag }(\hat n + 1) - \hat nf(\hat n){f^\dag }(\hat
       n).
\end{eqnarray}
On the other hand, with respect to the algebraic structure of Fokas-Lagerstrom superintegrable systems, Eq. (\ref{9}), we have
\begin{eqnarray}\label{17}
       [\hat A,{\hat A^\dag }] = \Phi (\hat H,N + 1) - \Phi (\hat H,N).
\end{eqnarray}
Now, if we deal with a constant energy, $E_{N}$, then $ \Phi ( \hat H,\hat n)$ depends only on $\hat n$ and if we now compare Eq. (\ref{16}) with Eq. (\ref{17}), we find that:
\begin{eqnarray}\label{18}
       {{n{f^2}(n)=\Phi ({E_N},n)}}.
\end{eqnarray}
Therefore, we can consider the algebra of Fokas-Lagerstrom system as a deformed Weyl-Heisenberg algebra with
the following deformation function:
\begin{eqnarray}\label{19}
      f(n) = \sqrt {16(N + 1 - n)(N + A - n)(N + B - n)}.
\end{eqnarray}
Now, by using the deformed creation and annihilation operators, $A$ and $A^\dag$, we arrive at
\begin{eqnarray}\label{20}
     \hat A|0\rangle =0= \hat A^\dag |N\rangle .
\end{eqnarray}
Thus, we conclude that for any constant $N$, corresponding to the constant value of energy $E_N$, there is a Hilbert space with finite dimension.\\
In the next section, we intend to construct the finite-dimensional coherent states corresponding to the Fokas-Lagrestrom
potential.

\section{Fokas-Lagrstrom nonlinear coherent states}\label{FLCS}
Let us now turn to define the finite-dimensional nonlinear coherent states corresponding to the Fokas-Lagrstrom potential. We follow the formalism of truncated coherent state approach introduced in~\cite{Kuang}, to define the Fokas-Lagrstrom nonlinear coherent states (FLNCSs). Therefore, we have:
      \begin{equation}\label{21}
      {|z \rangle }_{F.L} = C^{- \frac{1}{2}}\left(|z|^2\right)\exp(z{\hat A^\dag }) |0\rangle,
     \end{equation}
where $z$ is a complex number.
After some calculations, the above nonlinear coherent state can be recast into the following form:
\begin{eqnarray}\label{22}
      \left| z \right\rangle_{F.L}  = {C^{ - \frac{1}{2}}}\left( {{{\left| {z} \right|}^2}} \right)\sum\limits_{n = 0}^N
      {\frac{{{z^n}}}{{\sqrt    {\rho\left( n \right)} }}} \left| n \right\rangle,
\end{eqnarray}
where $\rho(n)$  is defined as:
\begin{eqnarray}\label{23}
      \rho (n) &=& {(\frac{1}{{16}})^n}\left[\frac{{\Gamma (n+1)\Gamma (N - n + 1)}}{{N!}}\right]
      \left[\frac{{\Gamma (N + A - n)}}{{\Gamma (N + A)}}\right]\n\\
       &&\times
      \left[\frac{{\Gamma (N + B - n)}}{{\Gamma (N + B)}}\right].
\end{eqnarray}
Here, $\Gamma$ is the gamma function, and the normalization constant $C$ is given by,
       \begin{equation}\label{24}
       C \left( {{{\left| {z} \right|}^2}} \right) = \sum\limits_{n = 0}^N {\frac{{{{\left| {z} \right|}^{\,2n}}}}{{\rho
       \left( n \right)}}}.
       \end{equation}

\subsection{Resolution of identity}
In this section, we intend to show that the constructed FLNCSs form an overcomplete set. In
 other words, we would show:
 \begin{eqnarray}\label{25}
        \int d^2 z | z \rangle W( | z |^2) \langle z |  = \sum\limits_{n = 0}^N  | n \rangle \langle n |= \mathbb{I}.
 \end{eqnarray}
The resolution of identity can be achieved by finding a measure function $W({\vert{z}\vert}^2)$. For this purpose, by
substituting $\left| z\right\rangle_{F.L} $  from the Eq.~(\ref{22}) into Eq.~(\ref{25}) we obtain,
\begin{eqnarray}\label{26}
       \int d^2 z  | z \rangle_{F.L} W( | z |^2) _{F.L}\langle z |  =\pi\sum_{n=0}^\infty  \frac{\left| n \right\rangle \left\langle {n} \right|}{{\rho \left( n \right)}}                           {\int_0^\infty  d( {{|z|^2}}){{|z|^{2n}}{\frac{W\left( {{|z|^2}} \right)}{{C\left( {{|z|^2}}\right)}}}}},
\end{eqnarray}
where we have used: $z = |z|{e^{i\theta }}$ and ${d^2}z = \frac{1}{2}d{(|z|^2)}d\theta $. Now, by using the change of variable
$x=|z|^{2}$ and considering $ w\left( x \right) = \pi\frac{W\left( {{|z|^2}} \right)}{{C\left( {{|z|^2}}\right)}}$, we should have:
\begin{equation}\label{27}
      \int_0^\infty dx\ {x^n}\ {w\left( x \right) = \rho \left( n \right)}.
\end{equation}
The above integral is called the moment problem and well-known mathematical methods such as Mellin
transformations can be used to solve it~\cite{Sixdeniers,Mathai}. From definition the of Meijers $G$-function, it follows that,
\begin{eqnarray}\label{28}
    \int dx\ x^{k - 1} G_{p,q}^{m,n}
    \left( \beta x \Big{|}
     \begin{array}{*{20}{c}}
      {a_1,\cdots,a_n,a_{n + 1},\cdots,a_p}\\
      {b_1,\cdots,b_m,b_{m + 1},\cdots,b_q}
     \end{array}
    \right) \n\\
    = \frac{1}{{{\beta ^k}}}\frac{{\prod _{j = 1}^m\Gamma ({b_j} + k)\prod _{j = 1}^n\Gamma (1 - {a_j} - k)}}{{\prod _{j    = m + 1}^q}\Gamma (1 - {b_j} - k){\prod _{j= n + 1}^p}\Gamma ({a_j} + k)}.
\end{eqnarray}
Therefore, by comparing Eqs.~(\ref{27}) and~(\ref{28}), we find that the weight function can be written as,
\begin{eqnarray}\label{29}
   w(x)
   &=&
   \frac{16}{N!\ \Gamma(N + A)\ \Gamma(N + B)}\n\\
   &  &\times
   G_{3,1}^{1,3}
   \left( 16 x \Big{|}
    \begin{array}{*{20}{c}}
     {-N-1,\  -N-A,\ -N-B}\\
     {0}
    \end{array}
   \right).
\end{eqnarray}
In this manner, it is seen that the FLNCSs satisfy the resolution of identity and
consequently form an overcomplete set.

\section{ The Physical Generation of FLNCSs }\label{Gen}

To generate an arbitrary but finite superposition of Fock states, a physical scheme has been proposed in Ref.~\cite{VK}. In this method, N two-level atoms which are prepared in a specific superposition of the excited state $\left|a\right\rangle$ and the ground state $\left| b \right\rangle $ interact with a resonant mode of radiation field in a cavity via the
Jaynes-Cummings Hamiltonian. The cavity field is initially in the vacuum state. By choosing atoms with appropriate initial states, it is possible to produce the desired state. In the resonator, the measurement of the internal state of the atom after it has passed through the cavity leaves the quantum states of field in a pure state. After the interaction of  $(K-1)$`th atom with cavity field, we make a measurement on the atomic state. If we find the atom in the
ground state $\left| b \right\rangle $, the state of the radiation field reads:
  \begin{equation}\label{31}
     \left| \varphi ^{(k-1)} \right\rangle= \sum\limits_{n}{ \varphi ^{(k-1)}_{n} }\left| n \right\rangle .
      \end{equation}
If the atom found in the excited state, our attempt to create the desired field state fails and we go back to vacuum
 state and start the procedure again.
Therefore, the state of the atom-field system after the $K$`th atom in the atomic state
 $|a\rangle+i\varepsilon_{k}|b\rangle$ exit the cavity is given by
      \begin{eqnarray}\label{32}
          |\Phi^{(k)} \rangle &=& \sum\limits_{n} \varphi^{(k-1)}_{n} C^{(k)}_{n} |n,a\rangle -iS^{(k)}_{n}| n+1,b
          \rangle \\
          &&+i\epsilon_{k}C^{(k)}_{n-1}| n,b \rangle +\epsilon_{k} S^{(k)}_{n-1}| n-1,a \rangle,\n
      \end{eqnarray}
where
     \begin{eqnarray}\label{33}
     C_n^k &= &\cos \left( {g{\tau _k}\sqrt {n + 1} } \right),\\
     S_n^k &=& \sin \left( {g{\tau _k}\sqrt {n + 1} } \right).
     \end{eqnarray}
Besides, the new coefficients ${\varphi_{n}}^{(k)}$ are given in terms of the ${\varphi_{n}}^{(k-1)}$ as
     \begin{equation}\label{34}
     \varphi _n^{\left( k \right)} = S_{n - 1}^{\left( k \right)}\varphi _{n - 1}^{\left( {k - 1} \right)} -
     {\varepsilon _k}C_{n - 1}^{\left( k        \right)}\varphi _n^{\left( {k - 1} \right)}.
     \end{equation}
We are now in a position to generate the finite-dimensional FLNCSs ${\left| z \right\rangle _{F.L}}$
based on this approach. For this purpose, we need to obtain that field combination state,
    \begin{equation}\label{35}
    \left| {{\varphi ^{(N - 1)}}} \right\rangle  = \sum\limits_{n = 0}^{N - 1} {{\varphi _n}^{(N - 1)}\left| n
    \right\rangle },
    \end{equation}
of $N$ number states such that after the $N$`th atom with the atomic superposition $\left| a
 \right\rangle+i\epsilon_{N}\left| b \right\rangle$ passed through the cavity and has been detected in the ground
state, the field in the cavity  become,
    \begin{equation}\label{36}
    {\left| z \right\rangle _{F.L}} = \sum\limits_{n = 0}^N {{d_n}} \left| n \right\rangle ,
    \end{equation}
where
     \begin{equation}\label{37}
     {d_n} = {C^{ - \frac{1}{2}\,}}\left( {\,{{\left| {z} \right|}^2}} \right)\frac{{{z^n}}}{{\sqrt {\rho \left( n
     \right)} }}.
     \end{equation}
By using Eq.~(\ref{34}), we get a set of $N+1$ equations which $N$ coefficients ${{\varphi _n}^{(N - 1)}}$ and the  parameter $\varepsilon_{N}$ can be obtained from it. The unknown coefficient ${{\varphi _n}^{(N - 1)}}$ is given by~\cite{VK}
 \begin{equation}\label{38}
\begin{array}{l}
\varphi _n^{( {N - 1})} = \sum\limits_{\upsilon  = 1}^{N - n} {\left[ {\prod\limits_{\mu  = n}^{n + \mu  - 2}
{\frac{{C_\mu ^{\left( N \right)}}}{{S_\mu ^{( N )}}}} } \right]} \frac{{{d_{n + \upsilon }}}}{{S_{n + \upsilon  - 1}^{(
N )}}}\,\varepsilon _N^{\upsilon  - 1}.
\end{array}
 \end{equation}
We also have the characteristic equation for $\varepsilon_{N}$ as
  \begin{equation}\label{39}
\begin{array}{l}
{d_0} + \sum\limits_{\upsilon  = 1}^N {\left[ {\prod\limits_{\mu  = 0}^{\upsilon  - 2} {\frac{{C_\mu ^{\left( N
\right)}}}{{S_\mu ^{\left( N \right)}}}} } \right]} \;\frac{{{d_\upsilon }}}{{S_{\upsilon  - 1}^{\left( N
\right)}}}\,\varepsilon _N^\upsilon = 0.
 \end{array}
 \end{equation}
We choose $\varepsilon_{N}$ as one of the $N$ roots of the Eq. (\ref{39}) which is a polynomial of degree $N$. In order to
obtain other parameters ${\varepsilon_{N - 1}},...,{\varepsilon _2}$ and ${\varepsilon _1}$ by a recurrence relation, we take $|\varphi^{(N - 1)} \rangle $ as a new desired state which can be generated by sending $N-1$ atoms
through the cavity. By following the same procedure, $N-1$ coefficient $\varphi^{(N - 2)}$ and the parameter
$\varepsilon_{N-1}$ are obtained. Repeating the calculations yields a sequence of complex numbers ${\varepsilon
_{1}},{\varepsilon _2},...,{\varepsilon _N}$ that defines the internal state of the $N$ injected atoms into the
cavity in order to generate the desired state ${\left| z \right\rangle _{FL}}$.

To be specific, we illustrate  step by step  generation of the FLNCSs with $N=2$ by using the space distribution $Q$-function~\cite{Scully}. The $Q$-function of a pure state $\vert\psi\rangle$ is defined as
 \begin{equation}\label{47}
  \begin{array}{l}
   Q\left( \alpha \right)=\dfrac{{\vert\langle\alpha\vert\psi\rangle\vert}^{2}}{\pi},
  \end{array}
 \end{equation}
where $\vert\alpha\rangle$ is the standard coherent state.

\begin{figure}[!ht]
  \centering
  \includegraphics[scale=0.5]{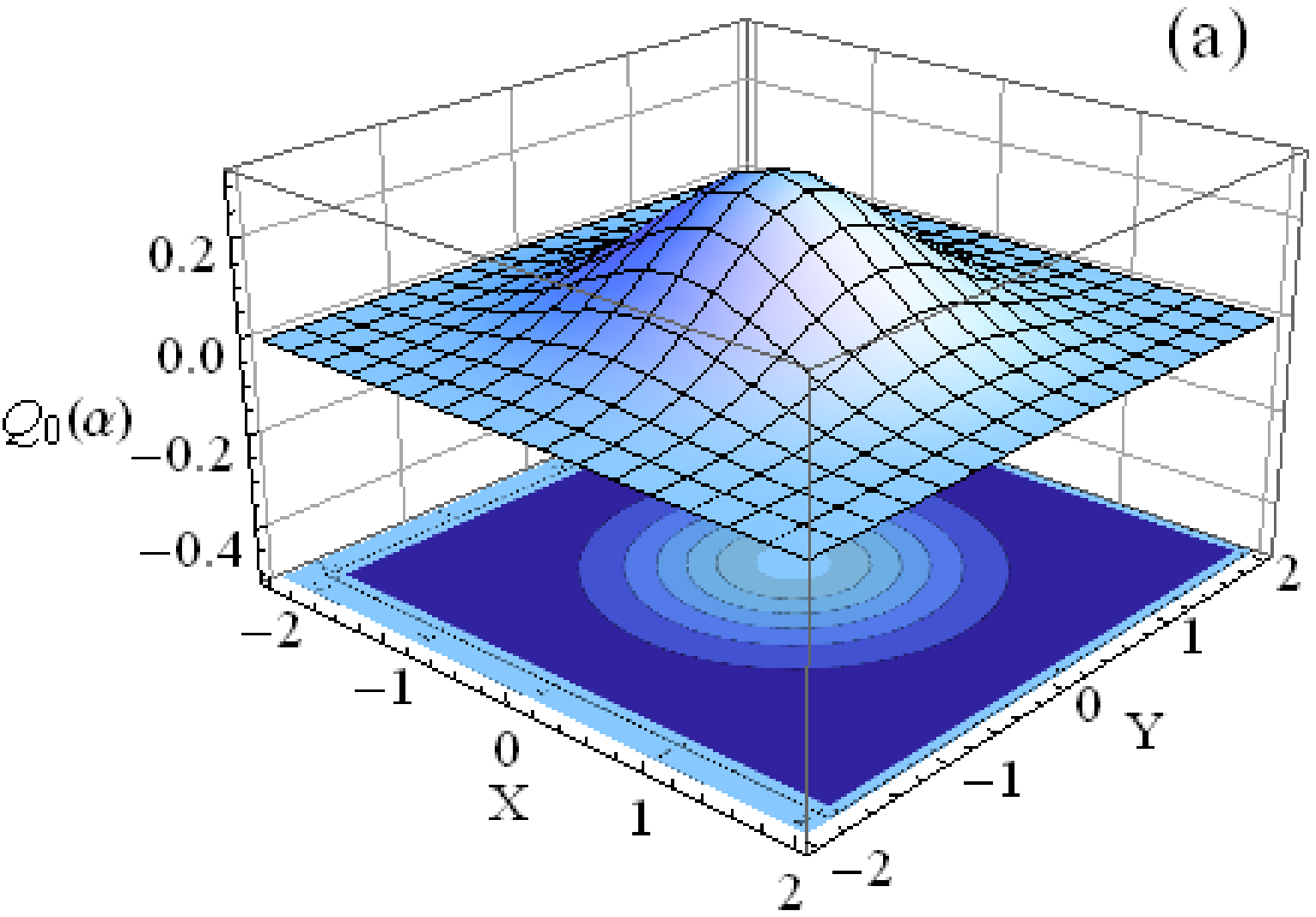}
  \includegraphics[scale=0.5]{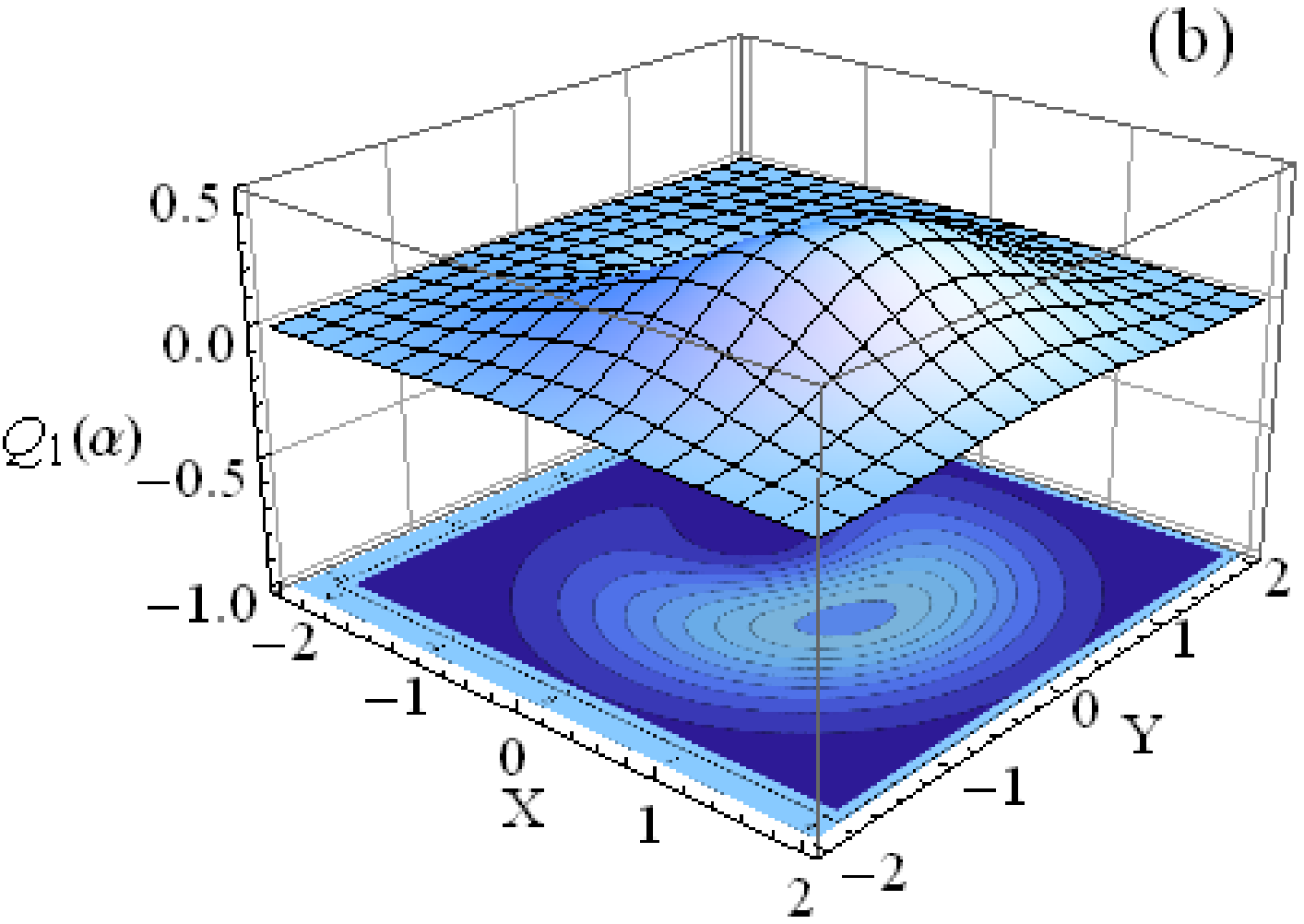}
  \includegraphics[scale=0.5]{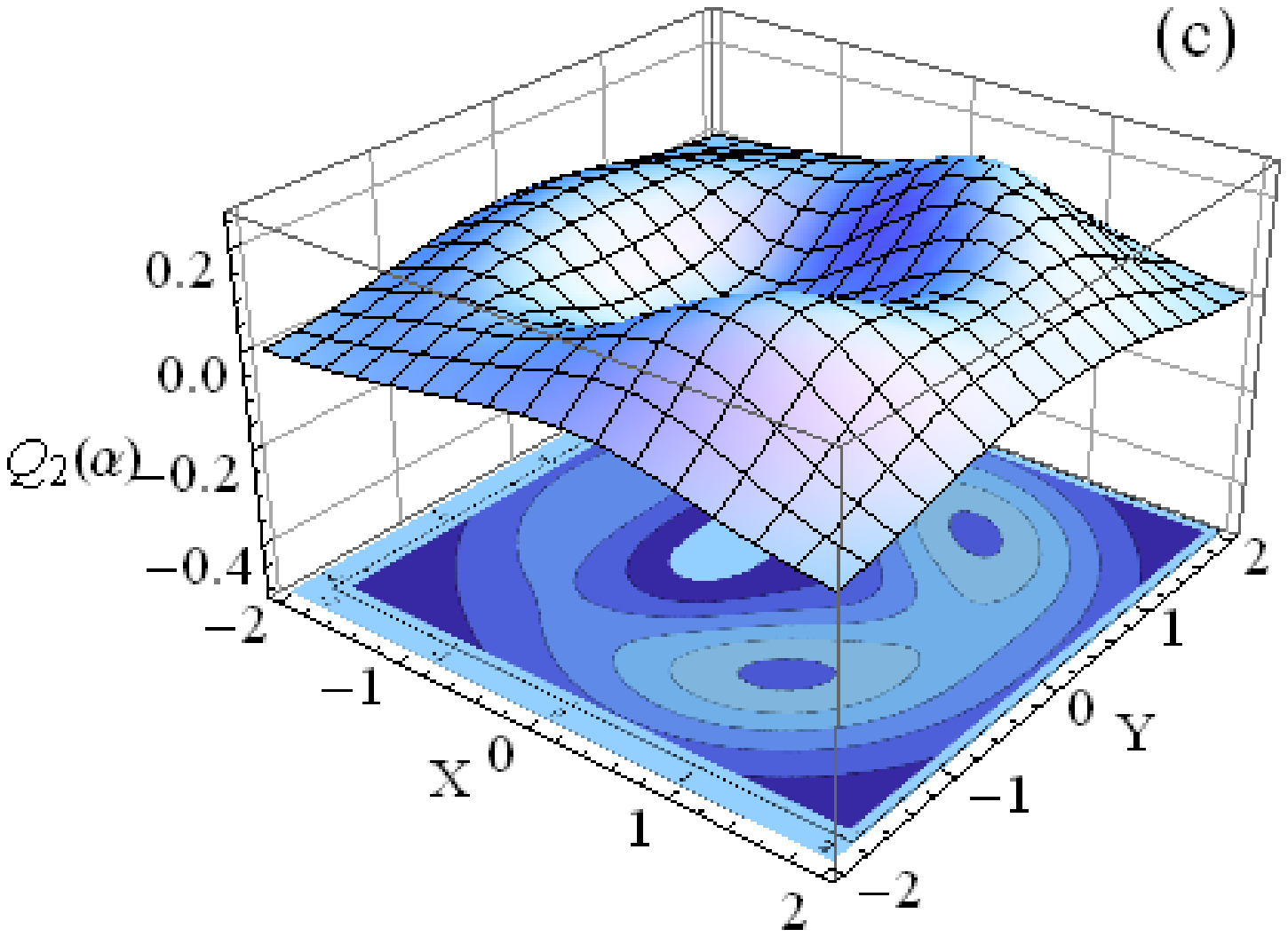}
  \caption{The $Q$-function $Q_{k}\left( \alpha \right) = \frac{\left|\langle\alpha|\varphi^{(k)}\rangle   \right|^{2}}{\pi}$ and the counter lines of the $Q$-function for the cavity state
 $\vert\varphi^{(k)}\rangle$ after the $k$`th atom has passed through the resonator and has been detected in ground
 state for a fix interaction parameter $g\tau=\dfrac{\pi}{5}$ and (a) $k=0$, (b) $k=1$ and (c) $k=2$ (the three steps for generation of the FLCSs with $N=2$).}
 \label{fig:d21}
\end{figure}

Fig.~\ref{fig:d21} displays the $Q$-function and the counter lines of the $Q$-function for the cavity state
 $\vert\varphi^{(k)}\rangle$ after the $k$`th atom has passed through the resonator and has been detected in ground
 state for a fix interaction parameter $g\tau=\dfrac{\pi}{5}$ and $k=0,1,2$.\\

%\begin{figure}[!ht]
% \centering
%  \includegraphics[scale=0.9]{d20}
%  \caption{\small{Contour line of the Q-function ${\color{blue}Q}(\alpha)={\vert\langle\alpha\vert\varphi^{1}\rangle\vert}^{2}/2$ for
%  the field state $\vert\varphi^{1}\rangle$ after the 1th atom has interacted with the field and has been detected in
%  the ground state. }}
% \label{fig:d20}
%\end{figure}

%\begin{figure}[!h]
%\centering
%\includegraphics[scale=0.9]{d22}
%\caption{\small{Contour line of the Q-function  ${\color{blue}Q}(\alpha)={\vert\langle\alpha\vert\varphi^{2}\rangle\vert}^{2}/2$ for
%the field state $\vert\varphi^{2}\rangle$ after the 2th atom has interacted with the field and has been detected in the
%ground state. }}
%\label{fig:d22}
%\end{figure}

\section{Quantum Statistical Properties of  the FLNCSs}\label{QStatistical}

In this section, we we shall proceed to study some quantum statistical properties of the FLNCSs,  including
probability of finding $n$ quanta, mean number of photons, Mandel parameter and quadrature squeezing.
\subsection{Photons-number distribution}

\begin{figure}[!ht]
  \centering
  \includegraphics[scale=0.6]{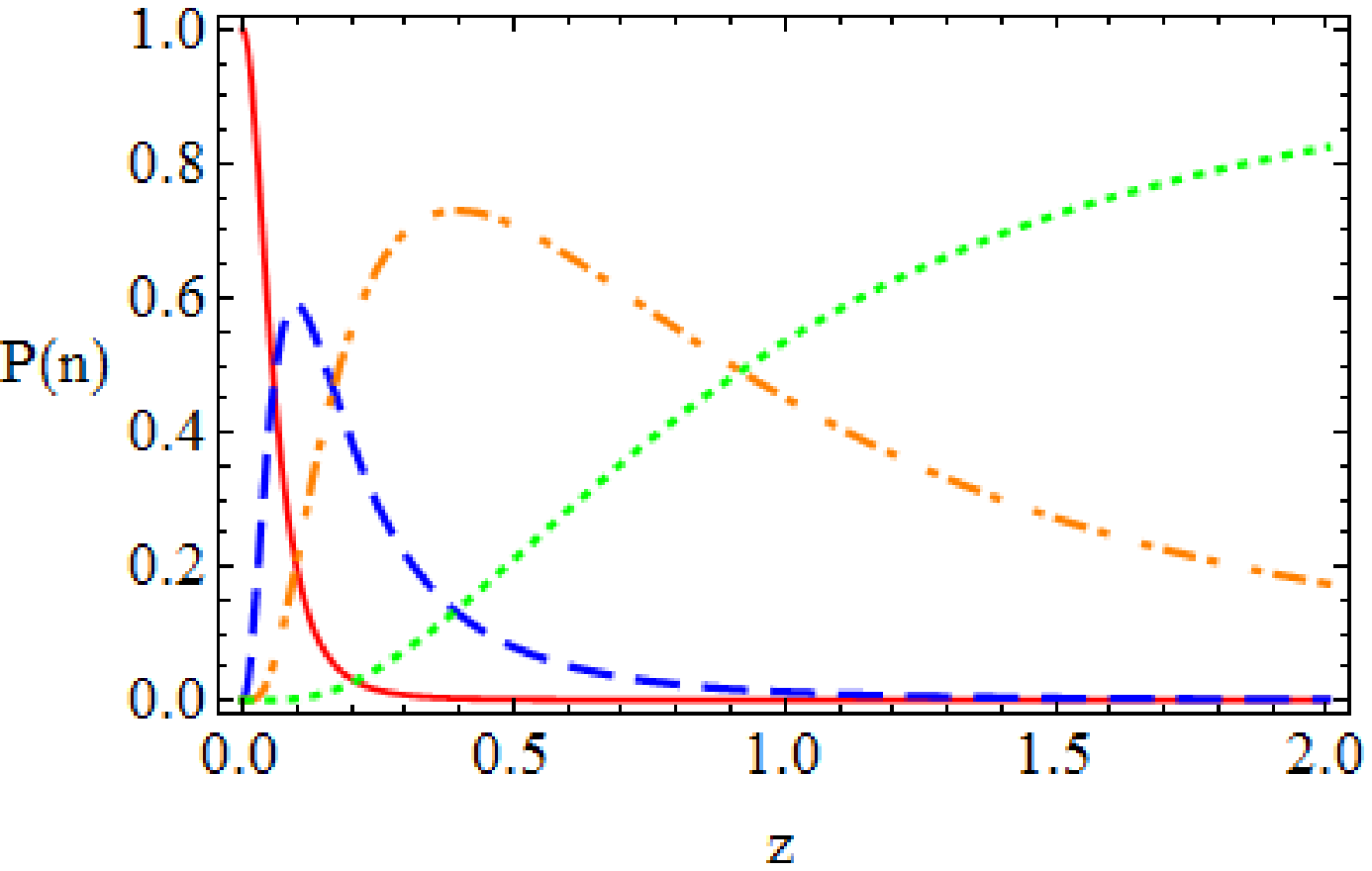}
  \caption{Probability of finding $n$ photons in the FLNCSs, versus $z$, for N=3, $A={2}/{3}$ and $B={1}/{3}$. Here, the solid red, dashed blue, dashed-dotted orange and dotted green lines correspond to $P(0)$, $P(1)$, $P(2)$ and $P(3)$, respectively. }
  \label{fig:d1}
\end{figure}

By using Eq. (\ref{22}), the probability $P(n)$ of finding $n$ photon in the FLNCSs is given by
       \begin{equation}\label{47}
       P(n) = \frac{1}{\Sigma_{n=0}^{N}\frac{|z|^{2n}}{\rho(n)}}\ \frac{|z|^{2n}}{\rho(n)}.
      \end{equation}
As it is clear from the above complex equation, it is difficult to predict the results analytically. Therefore, in Fig.~\ref{fig:d1} we show the effect of the parameter $z$ on the probability of finding $n$ photons in the FLNCSs with $A={2}/{3}$ and $B={1}/{3}$.
To get further insight, let us consider the limiting case $z\longrightarrow\infty$. Form Eq.~(\ref{47}), we obtain that the probabilities $P(n)$ tends to $\delta_{n,N}$. In other words, by increasing $z$, the FLNCS $\vert
z\rangle$ approaches to the number state $\vert N\rangle$.\\

\begin{figure}[!ht]
\centering
\includegraphics[scale=0.5]{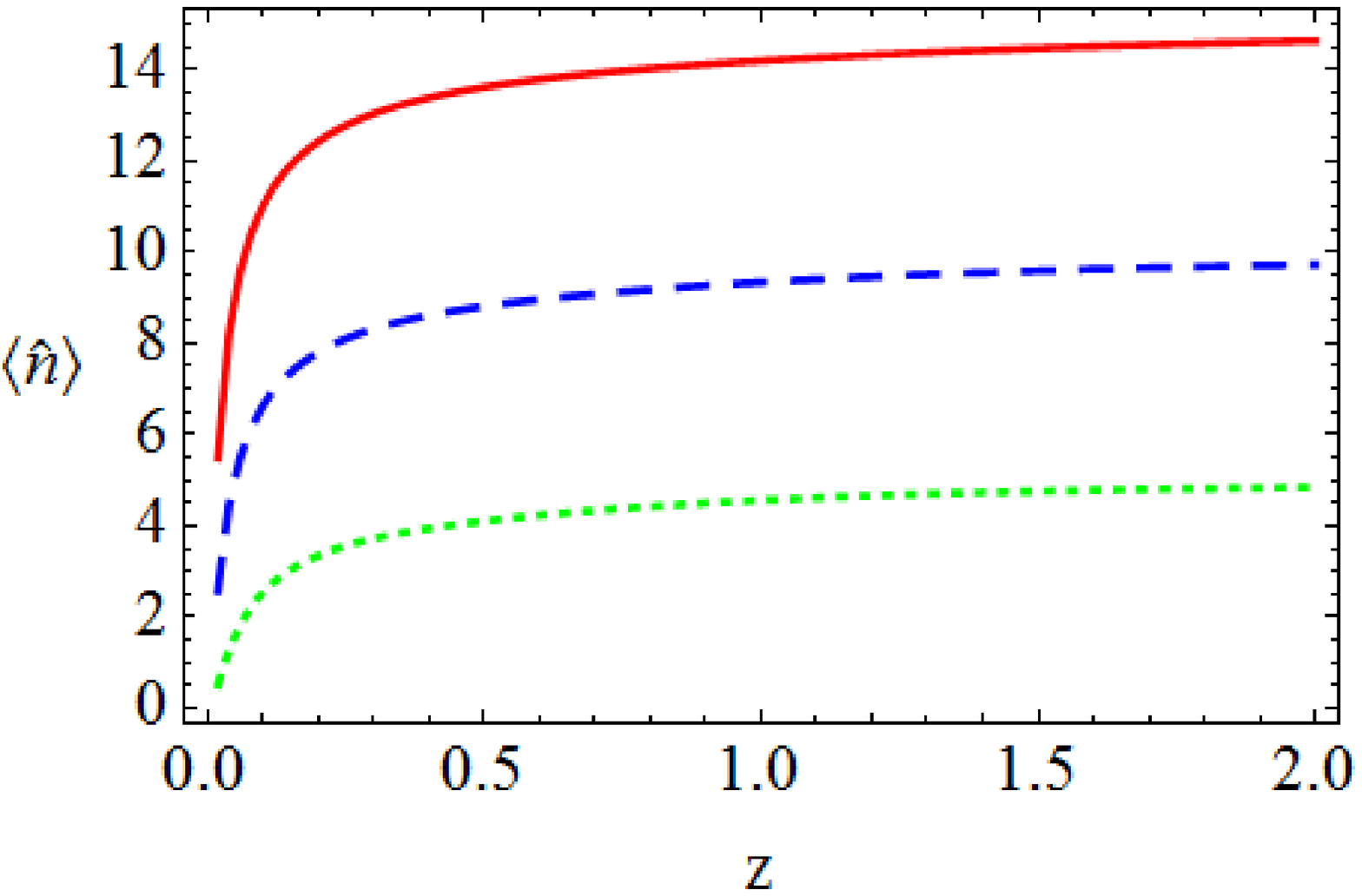}
\caption{\small The  mean number of  photons in  FLNCSs versus $z$ for $A={2}/{3}$ and
$B={1}/{3}$. Here, the dotted, dashed and solid lines correspond to $N=5$, $N=10$ and $N=15$, respectively.}
\label{fig:d2}
\end{figure}

The mean number of photons in the FLNCSs is calculated as follows:
   \begin{equation}\label{48}
         \left\langle \hat{n} \right\rangle =_{F.L}\left\langle z \right|{\hat a^\dag }\hat a\left| z \right\rangle_{F.L} =\sum_{n=0}^{N} n P(n).
    \end{equation}
In Fig. \ref{fig:d2}, the  mean number of  photons in  FLNCSs is plotted in terms of $z$ for $(A,B)=(2/3,1/3)$ and for different values of $N$.
It is seen that  for a constant $N$, by increasing $z$, the mean number of photons  increases, and in the limit of $z\longrightarrow\infty$ we get:  $_{F.L}\left\langle z \right|\hat  n\left| z \right\rangle_{F.L}  \to N$.
In addition, for a fixed value of $z$, the mean number of photons is increased by the  increasing the dimension of Hilbert space $N$. It is worth noting that these results are consistent with the results in  Fig. \ref{fig:d1}.

\subsection{Mandel parameter}

\begin{figure}[!ht]
\centering
\includegraphics[scale=0.55]{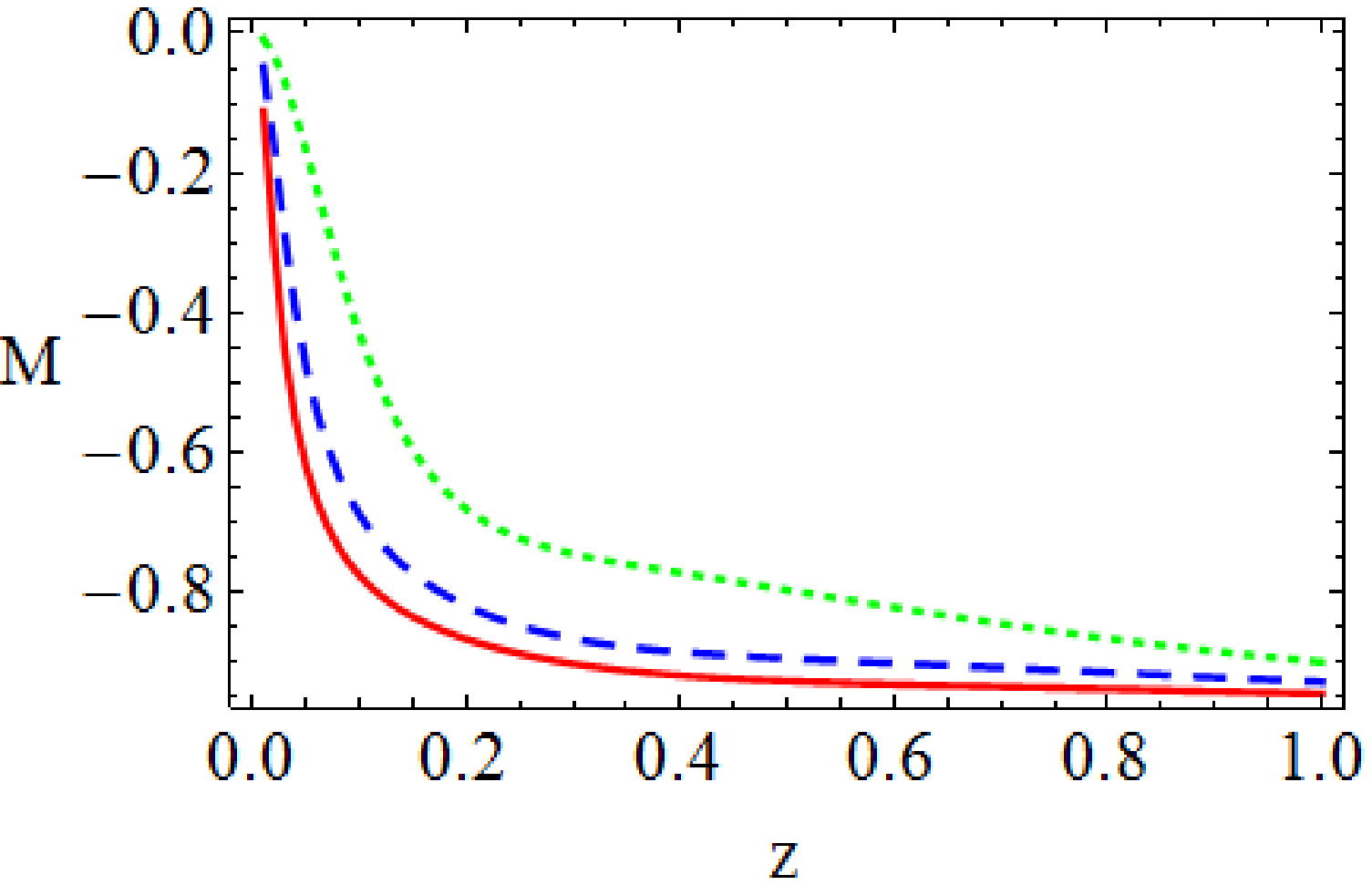}
\caption{Mandel parameter of the FLNCSs versus $z$ for $A={2}/{3}$ and
$B={1}/{3}$. Here, the dotted green line corresponds to N = 2, the dashed blue line
to N = 4 and the solid red line to N = 6.}
\label{fig:d5}
\end{figure}

In this subsection, we investigate deviation from the Poisson distribution for the  FLNCSs by using the Mandel parameter. This parameter is defined as~\cite{Mandel}:
 \begin{equation}\label{49}
   M = \frac{{{{(\Delta n)}^2} - \left\langle n \right\rangle }}{{\left\langle n \right\rangle }},
 \end{equation}
where the positive, zero and negative values represent super-Poissonian, Poissonian and sub-Poissonian distribution, respectively.
Due to the complexity of the final form of the Mandel parameter for the FLNCSs, we do not attempt to obtain its analytic form. Instead, we numerically study the Mandel parameter for these states. In Fig.~\ref{fig:d5}, we have plotted the Mandel parameter of the  FLNCSs
with respect to $z$ for $(A,B)=(2/3,1/3)$ and different values of $N$.
The results show that, for a fixed value of $N$, the photon counting statistic of the FLNCSs becomes more sub-Poissonian with increasing $z$ and then tends to $-1$ at very large values of z. We can justify this result by the fact that the quantum number state $\left| N \right\rangle$ has the Mandel parameter $M=-1$; since, the FLNCS approaches to the state $\left| N \right\rangle$ with increasing $z$, as is seen in Fig.~\ref{fig:d1}.

\subsection{Quadrature squeezing}

\begin{figure}[!ht]
\centering
\includegraphics[scale=0.5]{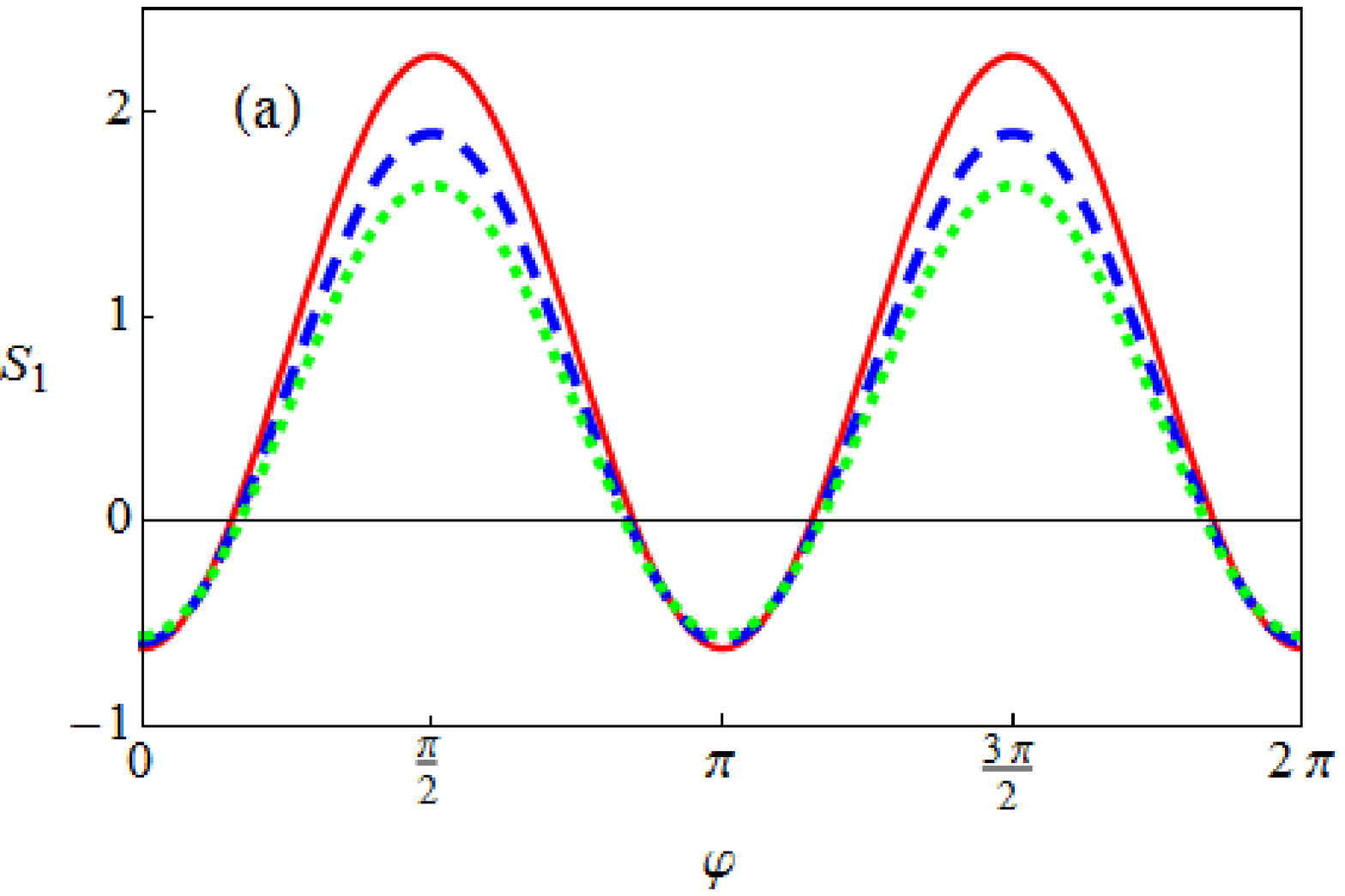}
\includegraphics[scale=0.5]{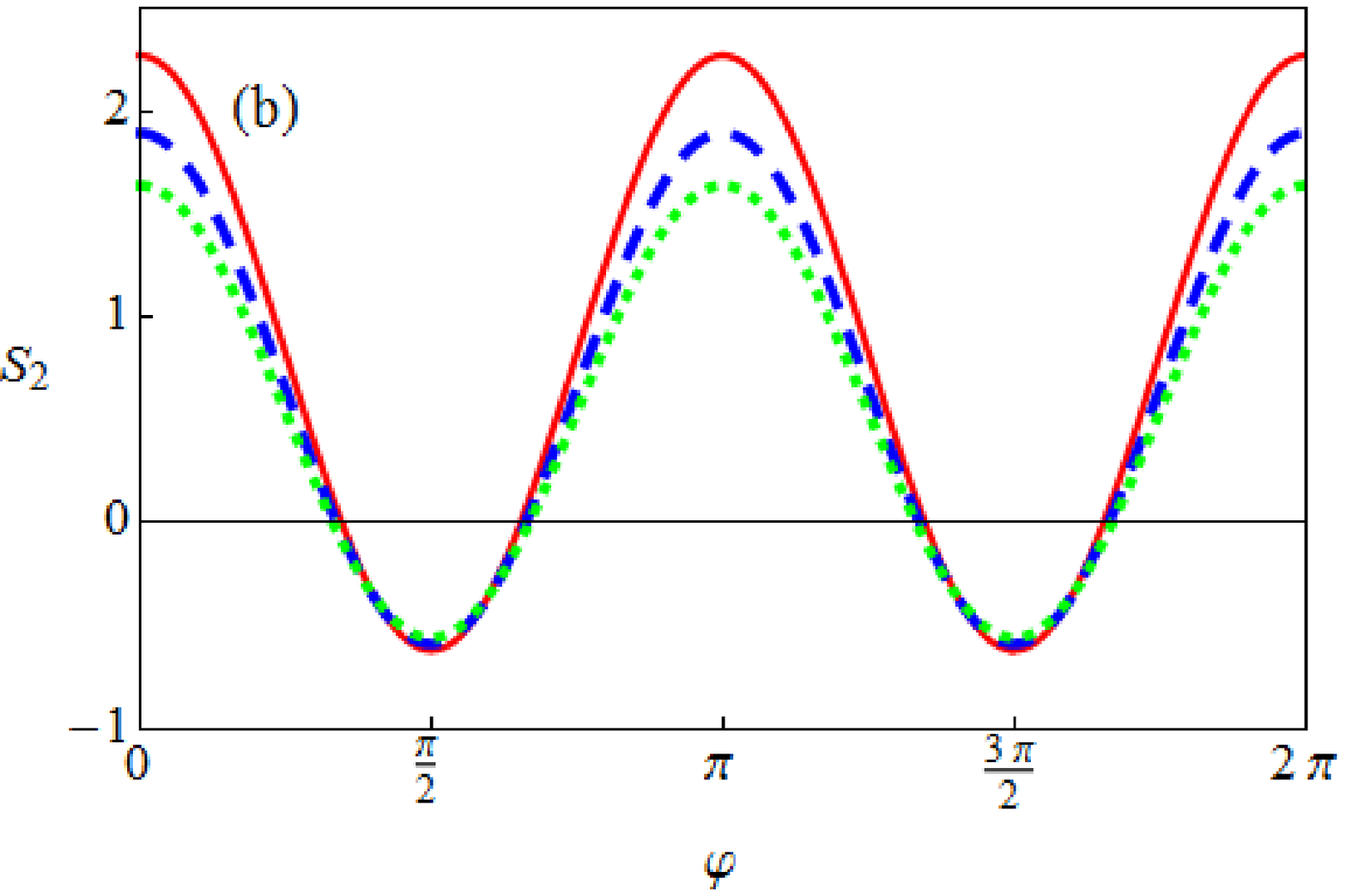}
\caption{The squeezing parameters (a) $S_{1}$ and (b) $S_{2}$ versus $\varphi$ for $z=0.02$  and $ (A,B) =({2}/{3},{1}/{3})$. Here, the
solid-red, dashed-blue and dotted-green lines correspond to $N=20$, $N=17$ and $N=15$, respectively.}
\label{fig:d8}
\end{figure}

In this subsection, we consider  the quadrature operators $\hat X_1$ and $\hat X_2$ defined in terms of creation and
annihilation operators  $\hat a$ and $\hat a^\dagger$ as follows :
 \begin{equation}\label{51}
   \begin{array}{l}
    {{\hat X}_1} = \frac{1}{2}\left( {\hat a{e^{i\phi }} + {{\hat a}^\dag }{e^{ - i\phi }}} \right),\\
    {{\hat X}_2} = \frac{1}{2i}\left( {\hat a{e^{i\phi }} - {{\hat a}^\dag }{e^{ - i\phi }}} \right).
   \end{array}
 \end{equation}
By using the commutation relation of $\hat a$ and $\hat a^\dagger$, the following uncertainty relation is obtained
\begin{equation}\label{52}
{\left( {\Delta {X_1}} \right)^2}{\left( {\Delta {X_2}} \right)^2} \ge \frac{1}{{16}}{\left| {\left\langle {\left[
{{{\hat X}_1},\left. {{{\hat X}_2}} \right]} \right.} \right\rangle } \right|^2} = \frac{1}{{16}}.
\end{equation}
As is known, the quadrature squeezing occurs if we have $ (\Delta X_i)^2<1/4 (i=1 or 2) $ or equally $S_i\equiv
4{(\Delta X_i)^2}-1<0 $.
% So for the squeezed state, at most one quadrature variance are reduced below the vacuum value $(\Delta X_{1})_{\vert0\rangle}^{2}=(\Delta X_{2})_{\vert0\rangle}^{2}=\dfrac{1}{4}$.
Figs.~\ref{fig:d8}(a) and~\ref{fig:d8}(b) display the squeezing parameters $S_{1}$ and $S_{2}$, respectively, for the FLNCSs with $(A,B)=(2/3,1/3)$ as a function of
$\varphi$ for $z=0.02$ and different values of $N$.

%\begin{figure}[!ht]
%\centering
%\includegraphics[scale=0.8]{d11}
%\caption{\small{Plot of squeezing parameter $S_{2}$ versus $\varphi$ for $z=0.02$  and $ (A,B) =({2}/{3},{1}/{3})$ the
%solid corresponds to $N=25$, the dashed to $N=20$, the dotted  to $N=15$} }
%\label{fig:d11}
%\end{figure}

It is seen that the range of the parameter $\phi$, in which the squeezing occurs, decreases by increasing $N$. In this range, the effect of increasing the dimension $N$ on the squeezing is dependant on $\phi$.
Besides, in the areas where the maximum squeezing occurs, an increase in $N$ leads to increase the squeezing,
while for other $\phi$'s available in this range, increasing of $N$ causes a reduction in the squeezing.\\

\section{Summary and Concluding Remarks}\label{Summary}

In this paper, we have introduced an algebraic approach to the Fokas-Lagerstrom system.
We have found that the two-dimensional Fokas-Lagerstrom algebra  can be considered as a deformed one-dimensional harmonic oscillator algebra. In this manner, we have succeeded to construct the nonlinear coherent states for this potential and studied
their quantum statistical properties. Finally, we have proposed a physical scheme to generate the FLNCSs.

%//////////////////////////////////////////////////////////////////////////////////////////

\end{document}